\documentclass[preprint,12pt,showpacs]{revtex4}

\usepackage[brazil, english]{babel}
\usepackage[utf8]{inputenc}
\usepackage{graphicx}
\usepackage{epsfig}
\usepackage{amssymb}
\usepackage{amsmath}
\usepackage{slashed}

\begin{document}

\title{Glueball spectra and Regge trajectories from a modified holographic softwall model}
\author{Eduardo Folco Capossoli$^{1,2,}$}
\email[Eletronic address:]{educapossoli@if.ufrj.br}
\author{Henrique Boschi-Filho$^{1,}$}
\email[Eletronic address: ]{boschi@if.ufrj.br}  
\affiliation{$^1$Instituto de F\'{\i}sica, Universidade Federal do Rio de Janeiro, 21.941-972 - Rio de Janeiro-RJ - Brazil \\
 $^2$Departamento de F\'{\i}sica, Col\'egio Pedro II, 20.921-903 - Rio de Janeiro-RJ - Brazil}

\begin{abstract}
In this work we propose a modified holographic softwall model, analytically solvable, to calculate the masses of lightest scalar glueball and its radial excitations and of higher spin glueball states for both even and odd spins. From these results  we obtain their respective Regge trajectories, associated with the pomeron for even spins and with the odderon for odd spins.  These results are in agreement with those calculated using other approaches.
\end{abstract}


\maketitle

\section{Introduction}

Since 1997 the AdS/CFT or Anti de Sitter/Conformal Field Theory correspondence \cite{Maldacena:1997re, Gubser:1998bc, Witten:1998qj, Witten:1998zw, Aharony:1999ti} provides new techniques and methodologies to deal with non-abelian gauge theories. 
The AdS/CFT correspondence relates a conformal supersymmetric Yang-Mills (SYM) theory with symmetry group $SU(N)$ for large $N$ ($N\rightarrow\infty$)  in a flat Minkowski spacetime with $3 + 1$ dimensions, with a $IIB$ superstring theory in a curved space $10$ dimensions, which is five dimensional anti de Sitter space times a five dimensional hypersphere, or simply, $AdS_5 \times S^5$.

Since the super Yang-Mills theory is a conformal field theory it can not be directly related to theories with mass or energy scales such as QCD. After breaking conveniently the conformal symmetry one can build phenomenological models that may describe some (non-perturbative) properties QCD approximately. The models constructed in this way are generically known AdS/QCD models. 

Some works have dealt with this issue \cite{Polchinski:2001tt, Polchinski:2002jw, BoschiFilho:2002ta, BoschiFilho:2002vd}. In these two last works, which introduced the idea of what is now called the hardwall model, a hard cutoff was introduced at a certain value $z_{max}$ of the holographic coordinate $z$ of the $AdS_5$ space and this space was reduced to just a slice in the region $0 \leq z  \leq z_{max}$.

Another holographic AdS/QCD model was proposed introducing a prescribed background dilatonic field to play the role of a soft cutoff instead of the AdS slice. This is known as the softwall model and  
was successful in describing vector mesons \cite{Karch:2006pv} and their Regge trajectories which are linear in contrast with the ones coming from the hardwall model. 
It was extended to describe light glueball states in \cite{Colangelo:2007pt}. 

An interesting modification of the softwall model is to impose that the dilatonic field became dynamical satisfying the Einstein equations in five dimensions. This dynamical softwall model has been used to describe the mass of the scalar glueball state and its radial excitations with good agreement with lattice data \cite{Li:2013oda}. This dynamical model does not have analytical solutions so one has to lean on numerical analysis.

In this work, we are going to consider a modified softwall model inspired on its dynamical version but which have analytical solutions. We apply this model to calculate glueball masses for the scalar case and its radial excitations, and high even and odd spins and construct their Regge trajectories associated with the pomeron and the odderon. Before going into this modified softwall model we start extending the original softwall model for higher spin glueballs. 


\section{Higher spin glueballs in the softwall model}

In order to describe higher spins in the softwall  (SW)  model we start with the following action
\begin{equation}\label{acao_soft1}
S =  \int d^5 x \sqrt{-g} e^{-\Phi(z)}\left[g^{mn} \partial_m{\cal G}\partial_n{\cal G} + M^2_{5} {\cal G}^2\right],
\end{equation}
\noindent where the field ${\cal G}$ is related to the scalar glueball state with mass $M_{5}$ in the $AdS_5$ space, defined by the metric: 
\begin{equation}\label{gs}
ds^2 = g_{mn} dx^m dx^n = \frac{R^2}{z^2}(dz^2 + \eta_{\mu \nu}dy^\mu dy^\nu) \,,
\end{equation}
\noindent where $m,n=0,1,2,3,4$, refer to five dimensional space, and $\mu,\nu=0,1,2,3$, refer to four dimensions with $\eta_{\mu \nu} = {\rm diag}(-1, 1, 1, 1)$. 
Here, the dilatonic field is prescribed as 
\begin{equation}\label{phi}
\Phi(z) = k z^2
\end{equation}
\noindent exactly as in the original softwall model \cite{Karch:2006pv}. Actually, the above action differs from the one presented in \cite{Colangelo:2007pt} to describe scalar glueballs by the presence of the mass term in five dimensions. This term is important here to include higher spin states as we discuss below. The corresponding equations of motion are:
\begin{equation}\label{emosw}
\partial_m[\sqrt{-g} \;  e^{-\Phi(z)} g^{mn} \partial_n {\cal G}] - \sqrt{-g} e^{-\Phi(z)} M^2_{5} {\cal G} = 0\,. 
\end{equation}
that can be written, after a convenient decomposition of the 5-d glueball wave function ${\cal G}(z,x^\mu)= v(z) \exp{i q_{\mu} x^{\mu}}$, where $v(z)=\psi(z) (z/R)^{3/2}\exp{\frac 12 (kz^2)}$, 
as ``Schroedinger-like'' equation 
\begin{equation}\label{eq_7}
- \psi''(z) + \left[ k^2 z^2 + \frac{15}{4z^2}  + 2k  +\left(\frac{R}{z} \right)^2 M^2_{5} \right]\psi(z) = \, -q^2 \, \psi(z)
\end{equation}
which has a well known solution:
\begin{equation}\label{eigenf}
\psi_n(z) = {\cal N}_n \;{z^{t(M_{5}) +\frac{1}{2}}}\; _1F_1(-n; t(M_{5})+1, kz^2)\exp\{-{kz^2}/{2}\}
\end{equation}
\noindent where ${\cal N}_n$ is a normalization constant, $t(M_{5}) = \sqrt{4 + R^2 M^2_{5} }$, and $_1F_1 (-n,a,x)$ is the Kummer confluent hypergeometric function. The corresponding ``eigenenergies'' $-q^2=-q_\mu q^\mu$ are identified with the 4-d glueball squared masses
\begin{equation}\label{kkss}
 m_n^2 = \left[ 4n +4 + 2\sqrt{4 +M^2_{5} R^2}\right] k;\;\;\;\; (n=0, 1, 2, \cdots). 
\end{equation}
It is known through the AdS/CFT correspondence how to relate the operator in the boundary theory with fields in the $AdS_{5} \times S^5$ space. The conformal dimension $\Delta$ of a boundary operator is given by:
\begin{equation}\label{dim_delta}
\Delta = 2 + \sqrt{4 + R^2 M^2_{5}}
\end{equation}
For a pure SYM theory defined on the boundary, one has that the scalar glueball state $0^{++}$ is represented by the operator ${\cal O}_4$, given by:
\begin{equation}\label{fmn}
{\cal O}_4 = Tr(F^2) = Tr(F^{\mu\nu}F_{\mu \nu})
\end{equation}
which has conformal dimension $\Delta = 4$. So, the lightest scalar glueball $0^{++}$ is dual to the fields with zero mass $(M^2_{5} = 0 )$ in the $AdS_5$ space, then Eq. (\ref{kkss}) becomes:
\begin{equation}\label{kkss1}
m_n^2 = \left[ 4n +8\right]k; \;\;\;\; (n=0, 1, 2, \cdots).
\end{equation}
This is the result found in  \cite{Colangelo:2007pt} that represents the equation for the Regge trajectory for the lightest scalar glueball $(n = 0)$ and its radial excitations $n=1, 2, \cdots$.

In the references \cite{BoschiFilho:2005yh} and \cite{Capossoli:2013kb} the masses of higher spin glueballs and the Regge trajectories related to the pomeron and the odderon were calculated using the holographic hardwall model following \cite{deTeramond:2005su}. The idea is to insert $J$ symmetrized covariant derivatives in  a given operator with spin $S$ so that the total angular momentum after the insertion is $S+J$. In the case of the operator ${\cal O}_4 = F^2$, one gets:
\begin{equation}\label{4+J}
{\cal O}_{4 + J} = FD_{\lbrace\mu1 \cdots} D_{\mu J \rbrace}F,
\end{equation}

\noindent with conformal dimension $\Delta = 4 + J$ and spin $J$. The reference \cite{BoschiFilho:2005yh} used this approach to calculate the masses of glueball states $0^{++}$, $2^{++}$, $4^{++}$, etc and to obtain the Regge trajectory for the pomeron in agreement with those found in the literature. 

Then, for even spin glueball states using the SW model after the insertion of symmetrized covariant derivatives, and using that $\Delta = 2 + \sqrt{4 + R^2 M^2_{5}}$ (eq. (\ref{dim_delta})), one has:
\begin{equation}
M^2_{5}R^2 = J(J+4)\,; \qquad ({\rm even}\, J)\,.
\end{equation}
\noindent Inserting this result in Eq. (\ref{kkss}), one gets:
\begin{equation}\label{kksspar}
m_n^2 = \left[ 4n + 4 + 2\sqrt{4 +J(J+4)}\right]k; \;\;\;\; (n=0, 1, 2, \cdots, {\rm even}\, J )\,,  
\end{equation}
and for the particular cases of non-excited states $(n=0)$, one has:
\begin{equation}\label{kkssparn}
m_n^2 = \left[ 4 + 2\sqrt{4 +J(J+4)}\right]k\,; \qquad ({\rm even}\, J).
\end{equation}

On the other side, for odd spin glueballs, following \cite{Capossoli:2013kb} , the operator ${\cal O}_6$ that describes the glueball state $1^{--}$ is given by
\begin{equation}
 {\cal O}_{6} =SymTr\left( {\tilde{F}_{\mu \nu}}F^2\right),
 \end{equation} 

\noindent and through insertion of symmetrized covariant derivatives one has
\begin{equation}
{\cal O}_{6 + J} = SymTr\left( {\tilde{F}_{\mu \nu}}F D_{\lbrace\mu1 \cdots} D_{\mu J \rbrace}F\right),
\end{equation}

\noindent with conformal dimension $\Delta = 6 + J$ and spin $1+J$. Following this approach in the hardwall model \cite{Capossoli:2013kb}, the masses of glueball states $1^{--}$, $3^{--}$, $5^{--}$, etc and the Regge trajectory for the odderon were obtained in agreement with those found in the literature.

Then, for the case of the odd spin glueballs states, as $\Delta = 2 + \sqrt{4 + R^2 M^2_{5}}$ (eq. (\ref{dim_delta})), one finds 
\begin{equation}
M^2_{5}R^2 = (J+6)(J+2)\,; \qquad ({\rm odd}\, J), 
\end{equation}

\noindent so that one can read for the non-excited odd spin glueball states $(n=0)$
\begin{equation}\label{kkssimpn}
m_n^2 = \left[ 4 + 2\sqrt{4 +(J+6)(J+2)}\right]k; \;\;\;\; ( {\rm odd}\, J ). 
\end{equation}

\noindent A discussion of these results together with a numerical analysis will be presented in section IV. 


\section{The modified softwall model}

In order to obtain the modified softwall model, let us start describing  briefly the dynamical softwall model discussed in \cite{Li:2013oda}.  
The $5D$ action for the graviton-dilaton coupling in the string frame is given by:
\begin{equation}\label{acao_corda}
S = \frac{1}{16 \pi G_5} \int d^5 x \sqrt{-g_s} \; e^{-2\Phi(z)} (R_s + 4 \partial_M \Phi \partial^M \Phi - V^s_G(\Phi))
\end{equation}
\noindent where $G_5$ is the Newton's constant in five dimensions, $g_s$ is the metric tensor in the $5-$dimensional space, $\Phi$ is the dilatonic field and $V^s_G$ is the dilatonic potential. All of these parameters are in the string frame. 
The metric tensor has the following form:
\begin{equation}\label{g_s}
ds^2 = g^s_{mn} dx^m dx^n = b^2_s(z)(dz^2 + \eta_{\mu \nu}dx^\mu dx^\nu); \; \; \;b_s(z) \equiv e^{A_s(z)}
\end{equation}
\noindent following the notation of the previous section. 

Performing a Weyl rescaling, from the string frame to the Einstein frame, one can obtain the equations of motion for the dilaton $\Phi(z)$ and the metric represented by the function $A_s(z)$ which is a set of  coupled differential equations. 
Going back to the string frame and choosing $\Phi(z)=kz^2$, as in the original softwall model, one has the solutions  (see the Appendix~A):
\begin{equation}\label{redef_2}
 A_s(z) = \log{\left( \frac{R}{z} \right)}  + \frac{2}{3}\Phi(z) - \log{\left[_0F_1\left(\frac 54, \frac{\Phi^2}{9}\right)\right]}\,, 
\end{equation}
\noindent which means that the metric (\ref{g_s}) is a deformed AdS space and
\begin{equation}\label{vs}
 V^s_G(\Phi) =\exp\{-\frac 43 \Phi\} \left[ -\frac{12 ~ _0F_1(1/4, \frac{\Phi^2}{9})^2}{R^2} + \frac{16 ~ _0F_1(5/4, \frac{\Phi^2}{9})^2 \Phi^2}{3 R^2}\right]
\end{equation}
so that this potential generates the desired dilaton. 

Let us now describe the scalar glueball in 5D with the action in the string frame exactly as in Eq. (\ref{acao_soft1}) but with the metric replaced by (\ref{g_s}), and the corresponding equations of motion are:
\begin{equation}\label{eom_1}
\partial_M[\sqrt{-g_s} \;  e^{-\Phi(z)} g^{MN} \partial_N {\cal G}] - \sqrt{-g_s} e^{-\Phi(z)} M^2_{5} {\cal G} = 0\,.
\end{equation}
 One can solve the equations of motion using again the ansatz 
\begin{equation}\label{ansatz}
{\cal G}(z, x^{\mu}) = v(z) e^{i q_{\mu} x^{\mu}}\,,
\end{equation}
\noindent and defining $v(z) = \psi (z) e^{{B(z)} / {2}} $ where 
\begin{equation}\label{subs}
B(z) = \Phi(z) - 3A_s(z) \,,
\end{equation}
so that one gets a Schroedinger like equation: 
\begin{equation}\label{equ_5}
- \psi''(z) + \left[ \frac{B'^2(z)}{4}  - \frac{B''(z)}{2} + M^2_{5} \left( \frac{R}{z}\right)^2  e^{4kz^2/3} {\cal A}^{-2} \right] \psi(z) 
= - q^2 \psi(z)\,,
\end{equation}
where ${\cal A}= ~_0F_1(5/4, {\Phi^2}/{9})$. This equation was solved numerically in \cite{Li:2013oda}.

Inspired by this dynamical model, and seeking for analytical solutions, we propose a modified softwall model whose action is given by Eq. (\ref{acao_soft1}), with metric given by (\ref{g_s}) and the dilaton $\Phi(z)$ still given by (\ref{phi}) but with the function 
$A_s(z)$ replaced by: 
\begin{equation}\label{am}
{{A}}^s_M(z) = \log{\left( \frac{R}{z} \right)}  + \frac{2}{3}\Phi(z). 
\end{equation}
Looking at (\ref{g_s}) and (\ref{am}) one can note that this modified softwall model is no longer $AdS_5$.
This is also true for the dynamical softwall model.  But for $z\rightarrow 0$ which means the UV limit in both cases, it can be seen that $  A_s(z)|_{z\rightarrow 0} \to {{A}}^s_M(z)|_{z\rightarrow 0} \propto \log \left( \frac{R}{z} \right)$. This means that the geometry still remains $AdS_5$ in the UV limit when $A_s(z)$ is replaced by $A^s_M(z)$.

Then eq.(\ref{equ_5}) can be read as:
\begin{equation}\label{equ_7}
- \psi''(z) + \left[ k^2 z^2 + \frac{15}{4z^2}  - 2k + M^2_{5} \left( \frac{R}{z}\right)^2  e^{4kz^2/3}\right] \psi(z) = (- q^2 )\psi(z).
\end{equation}
This equation is a Schroedinger-like equation with effective potential given by 
$${\cal V}(z) = \left[ k^2 z^2 + \frac{15}{4z^2} - 2k + M^2_{5} \left( \frac{R}{z}\right)^2  e^{4kz^2/3} \right]\,.$$
This is still not exactly solvable so we expand the exponential in the last term in the brackets and just retain terms up to first order in the parameter $k$ \cite{comment}. This procedure gives us the equation 
\begin{equation}\label{equ_7_1_new}
- \psi''(z) + \left[ k^2 z^2 + \frac{15}{4z^2}  - 2k + M^2_{5} \left( \frac{R}{z}\right)^2   + \frac{4 kz^2}{3}M^2_{5} \left( \frac{R}{z}\right)^2\right] \psi(z) = (- q^2 )\psi(z)\,,
\end{equation}
which is exactly solvable and represents the modified softwall model that we consider here,
which can also be written as 
\begin{equation}\label{eq_7_2}
- \psi''(u) + \left[ u^2 + \frac{t^2 - \frac{1}{4}}{u^2} \right]\psi(u) = \left[ \frac{- q^2}{k}+2 -\frac{4}{3}R^2M^2_{5}\right] \psi(u)\,,
\end{equation}

\noindent where  $u=\sqrt{k} \, z^2$ and $t = \sqrt{4 + R^2 M^2_{5} }$. 
From the eigenenergies  and associating $-q^ 2_n$ with the square of the masses of the 4D glueball states, one has:
\begin{equation}\label{adsw_1}
m_n^2 = \left[ 4n + 2\sqrt{4 +M^2_{5} R^2} + \frac{4}{3}R^2M^2_{5} \right]k; \;\;\;\; (n=0, 1, 2, \cdots). 
\end{equation}
and the eigenfuntions are still given by (\ref{eigenf}). 

So, for the lightest scalar glueball $0^{++}$  dual to the fields with zero mass $(M^2_{5} = 0 )$ in the $AdS_5$ space, the Eq. (\ref{adsw_1}) becomes:
\begin{equation}\label{adsw}
m_n^2 = \left[ 4n +4 \right] k \,.
\end{equation}

For even spin glueball states we have $M^2_{5}R^2 = J(J+4)$ as in our previous discussion on the original softwall model and just computing the masses for non excited states $(n=0)$, one gets:
\begin{equation}\label{adsw_1_even}
m_n^2 = \left[  2\sqrt{4 +J(J+4)} + \frac{4}{3}J(J+4) \right]k\,; \qquad ({\rm even}\, J).
\end{equation}

For odd spins glueball states, with $M^2_{5}R^2 = (J+6)(J+2)$, one has
\begin{equation}\label{adsw_1_odd}
m_n^2 = \left[  2\sqrt{4 +(J+6)(J+2)} + \frac{4}{3}(J+6)(J+2) \right]k\,; \qquad ({\rm odd}\, J).
\end{equation}

A comparison between these results and the ones from the original SW model will be presented in the next section. 


\section{Numerical analysis}

Now we are going to obtain numerical values for the various masses discussed in this work. 
For comparison, we show in Table \ref{t2} the values of the masses for the scalar glueball and its excitations calculated from the lattice. 

\begin{table}[h]
\centering
\begin{tabular}{|c|c|c|c|c|c|}
\hline 
  & ref. \cite{Meyer:2004gx} & ref. \cite{Morningstar:1999rf}  & ref. \cite{Chen:2005mg} & \multicolumn{2}{c|}{ref. \cite{Lucini:2001ej}} \\ 
\hline 
$J^{PC}$ & $N_c = 3$ & \multicolumn{2}{c|}{$N_c = 3$ anisotropic lattice} & $N_c = 3$ & $N_c \rightarrow \infty$ \\ 
\hline 
$0^{++}$ & 1.475(30)(65) &1.730(50)(80) &1.710(50)(80) & 1.58(11) & 1.48(07) \\ 
\hline 
$0^{++}*$ & 2.755(70)(120) & 2.670(180)(130) &  & 2.75(35) & 2.83(22) \\ 
\hline 
$0^{++**}$ & 3.370(100)(150) &  &  &  &  \\ 
\hline 
$0^{++***}$ & 3.990(210)(180) &  &  &  &  \\ 
\hline 
\end{tabular} 
\caption{\em Lightest scalar glueball and its radial excitation masses expressed in GeV from lattice.}
\label{t2}
\end{table}
\begin{table}[h]
\centering
\begin{tabular}{|c|c|c|c|c|c|c|c|}
\hline
 &  \multicolumn{4}{c|}{Glueball States $J^{PC}$} &   \\  
\cline{2-5}
 & $0^{++}$ & $0^{++*} $ & $0^{++**}$ & $0^{++***}$  & $k$ \\
 \hline 
 $n$ & 0 & 1 & 2 & 3 & \\
\hline \hline
\, $m_n$ \, &\, 1.72 \, & \, 2.11 \, & \, 2.43 \, & \, 2.72 \, & \, 0.37 \, \\ \hline
\, $m_n$ \, & \, 2.32 \, & \, 2.83 \, & \, 3.27 \, & \, 3.66 \, & \, 0.67 \, \\ \hline 
\, $m_n$ \, &\, 2.53 \, & \, 3.10 \, &\, 3.58 \,& \, 4.00 \, & \, 0.80 \, \\ \hline 
\, $m_n$ \,  &\, 2.83 \, &\, 3.46 \,&\, 4.00 \,& \, 4.47 \, &  \,1.00 \, \\ \hline
\end{tabular}
\caption{\em Masses $m_n$ expressed in GeV for the glueball states $J^{PC}$ of the the lightest scalar glueball $(n=0)$ and its radial excitations $(n=1,2,3)$ from the original SW, using the Eq. (\ref{kkss1}) for various values of $k$ from $0.37$ to $1.00$ GeV$^2$.}
\label{t1}
\end{table}
\begin{table}[h]
\centering
\begin{tabular}{|c|c|c|c|c|c|c|c}
\hline
 &  \multicolumn{4}{c|}{Glueball States $J^{PC}$}  & \\  
\cline{2-5}
 & $0^{++}$ & $0^{++*} $ & $0^{++**}$ & $0^{++***}$ & $k$ \\ \hline
 $n$ & 0 & 1 & 2 & 3 & \\ 
\hline \hline
\, $m_n$ \,                                   
&\, 0.89\, &\, 1.26 \,&\, 1.55 \,& \, 1.79 \, & \, 0.20 \, \\ \hline
\, $m_n$ \,                                   
&\, 1.84\, &\, 2.61 \,&\, 3.19 \,& \, 3.69 \, & \, 0.85 \, \\ \hline
\, $m_n$ \,                                   
&\, 2.00\, &\, 2.83 \,&\, 3.46 \,& \, 4.00 \, & \, 1.00 \, \\ \hline
\end{tabular}
\caption{\em Masses expressed in GeV for the glueball states $J^{PC}$ of the the lightest scalar glueball and its radial excitations from the modified softwall model using Eq.(\ref{adsw}) for $k= 0.2,  \;0.85$ and 1~GeV$^2$.}
\label{t3}
\end{table}

 Let us start with the predictions for the scalar $0^{++}$ state. We begin with the result of Eq. (\ref{kkss1}) from the original SW model that represents the equation for the Regge trajectory for the lightest scalar glueball $(n = 0)$ and its radial excitations $(n=1, 2, \cdots)$. Calculating these masses for various values of $k$ from $0.37$ to $1.00$ GeV$^2$ one gets the results shown in Table  \ref{t1}. 
Comparing these values with the ones shown in Table \ref{t2}, one sees that in general these masses do not fit those from the lattice. However, note that for $k=0.67$ GeV$^2$ one fits the masses of the three excited states $n=1,2,3$, but not the ground state $n=0$.

On the other hand, the masses derived from the Regge trajectory (\ref{adsw}) using the modified SW model and $k=0.2, \;0.85$ and $ 1$ GeV$^2$ are presented in the table \ref{t3}.
Note that for  $k=0.85$  GeV$^2$ the agreement with lattice is good.  

Now, let us move to the case of high even spins. The masses found from Eq. (\ref{kkssparn})  in the original softwall model for higher spins with even $J$ and $k= 1$ and 2 GeV$^2$ are shown in the table  \ref{t5}. From the results with  $k= 2$  GeV$^2$  one can derive the Regge trajectory for even glueball states associated with the pomeron:
\begin{equation}\label{rtpsw}
J(m^2) = 0.25 m^2 - 4\,,
\end{equation}
\noindent where $J$ is the glueball state spin and $m^2$ is the glueball state mass squared. 
This Regge trajectory has a good slope but the intercept is not in agreement with the literature \cite{Landshoff:2001pp}.

From the modified SW model, the masses found from Eq. (\ref{adsw_1_even}) for higher spins with even $J$ and $k= 0.2$ GeV$^2$ are also shown in the table  \ref{t5}. From these results one can derive the Regge trajectory for even glueball states which can be associated with the pomeron:
\begin{equation}\label{rtpadsw}
J(m^2) = (0.23 \pm 0.02) m^2 + (0.82 \pm 0.51)\,.
\end{equation}
\noindent The errors for the slope and the intercept come from the linear fit. This Regge trajectory is in agreement with that presented for the pomeron  \cite{Landshoff:2001pp}.

\begin{table}
\centering
\begin{tabular}{|c|c|c|c|c|c|c|c|}
\hline
 &  \multicolumn{6}{c|}{Glueball States $J^{PC}$}  & \\  
\cline{2-7}
 & $0^{++}$ & $2^{++} $ & $4^{++}$ & $6^{++}$ & $8^{++}$ & $10^{++}$  & $ k $ \\
\hline \hline
Masses                                   
&\, 2.83\, &\, 3.46 \,&\, 4.00 \,& \, 4.47 \, &\, 4.90&\, 5.29 \, & \,  1.00 \,   \\ \hline 
Masses                                   
&\, 4.00    \, &\, 4.90  \,&\, 5.67    \,& \, 6.32     \, &\, 6.93   \, &\,  7.48   \, & \,  2.00 \,   \\ \hline 
Masses                                   
&\, 0.89\, &\, 2.19 \,&\, 3.30 \,& \, 4.38 \, &\, 5.44 &\, 6.49 \, & \, 0.20 \, \\ \hline 
\end{tabular} 
\caption{\em Masses expressed in GeV for the glueball states $J^{PC}$ with even $J$ from the original SW using Eq. (\ref{kkssparn}) with $k= 1$ and 2 GeV$^2$ and from the modified SW using Eq. (\ref{adsw_1_even}) with $k= 0.2$ GeV$^2$.}
\label{t5}
\end{table}

A last comment about even glueball states:  one can choose another set of states, for exemple, $2^{++}, 4^{++}, 6^{++}, 8^{++}$, from Table \ref{t5} with $k$ = 0.20 GeV$^2$, and find the following Regge trajectory:
\begin{equation}\label{rtpadsw_0_10}
J(m^2) = (0.24 \pm 0.02) m^2 + (1.15 \pm 0.36)\,,
\end{equation}
\noindent which is still compatible with \cite{Landshoff:2001pp} and \cite{Meyer:2004jc} where it was argued that the state $0^{++}$ does not belong to the pomeron's Regge trajectory.

\begin{table}[h]

\centering
\begin{tabular}{|c|c|c|c|c|c|c|c|}
\hline
 &  \multicolumn{6}{c|}{Glueball States $J^{PC}$}  & \\  
\cline{2-7}
 & $1^{--}$ & $3^{--} $ & $5^{--}$ & $7^{--}$ & $9^{--}$ & $11^{--}$  & $ k $ \\
\hline \hline
Masses                                   
&\, 3.74\, &\, 4.24 \,&\, 4.69 \,& \, 5.10 \, &\, 5.48&\, 5.83\, & \, 1.00 \,  \\ \hline
Masses                                   
&\, 5.29   \, &\, 6.00    \,&\, 6.63    \,& \, 7.21   \, &\, 7.75    &\,  8.24   \, & \, 2.00 \,  \\ \hline

Masses                                   
&\, 2.82\, &\, 3.94 \,&\, 5.03 \,& \, 6.11 \, &\, 7.19&\, 8.26\, & \,  0.20 \, \\ \hline
\end{tabular}
\caption{\em Masses expressed in GeV for the glueball states $J^{PC}$ with odd $J$ from SW using eq.(\ref{kkssimpn}) and $k= 1$ and 2 GeV$^2$ and from the modified SW using eq.(\ref{adsw_1_odd}) and $k= 0.2$ GeV$^2$.} 
\label{t6}
\end{table}

Let us now discuss the case of odd spins. The masses found from Eq. (\ref{kkssimpn}) for the original SW model for higher odd $J$ spins with $k= 1$ and 2 GeV$^2$ are shown in the Table  \ref{t6}. 
From the results for $k$=2 GeV$^2$ one can derive the Regge trajectories for odd glueball states associated with the odderon:
\begin{equation}\label{rtosw}
J(m^2) = 0.25 m^2 - 6\,.
\end{equation}
\noindent 
This Regge trajectory is not in agreement with the ones presented in \cite{LlanesEstrada:2005jf}. 

The masses found from the modified softwall model for higher odd spins, given by Eq.  (\ref{adsw_1_odd}), with $k= 0.2$ GeV$^2$ are also shown in the Table  \ref{t6}. 
From these results one can derive the Regge trajectory for odd spin glueball states associated with the odderon:
\begin{equation}\label{rtoadsw}
J(m^2) = (0.17 \pm 0.01) m^2 + (0.40 \pm 0.44)\;.
\end{equation}
\noindent The errors for the slope and intercept come from the linear fit. This Regge trajectory for the odderon is in agreement with that presented in \cite{LlanesEstrada:2005jf}, within the nonrelativistic constituent model. 
%
One can also choose another set of odd glueball states, for example, $1^{--}, 3^{--}, 5^{--}, 7^{--},  9^{--}$ with $k=0.2$ GeV$^2$ and find the following Regge trajectory:
\begin{equation}\label{rtoadsw_1_9}
J(m^2) = (0.18 \pm 0.01) m^2 + (0.02 \pm 0.40),
\end{equation}
\noindent which is also compatible with \cite{LlanesEstrada:2005jf} within the nonrelativistic constituent model. In ref. \cite{LlanesEstrada:2005jf} it was argued that the odd spin glueball state $1^{--}$ might not belong to the Regge Trajectory associated with the odderon. In contrast, in this work, all Regge trajectories associated with the odderon contained the odd spin glueball state $1^{--}$.

At this point, one can note that the value chosen for the free parameter $k$ is not the same for scalar glueball and the higher spin glueballs. For the scalar glueball state and its radial excitations, the value of $k$ that provided a good Regge trajectory was $k= 0.85$ GeV$^2$. On other hand, for higher spin glueball states, the value of $k$ that provided good Regge trajectories was $k= 0.20$ GeV$^2$.
This difference points to a limitation of the modified softwall model but seems to be acceptable 
since we are dealing with a phenomenological model.


\section{Conclusions}

In this work we used first the original softwall model to describe high spin states glueballs and obtained 
not soo good results. Then we proposed a modified softwall model inspired in a dynamical model presented in \cite{Li:2013oda}. From this modified softwall model we obtained good results for the Regge trajectories of the scalar glueball state and its radial excitations and the Regge trajectories for the pomeron and the odderon, in good agreement with the literature 
\cite{Meyer:2004gx,Morningstar:1999rf,Chen:2005mg,Lucini:2001ej,Landshoff:2001pp,LlanesEstrada:2005jf,Meyer:2004jc}.

An important thing to be commented about the even spin glueball states. Due to the fact that in this work we use the free parameter $k = 0.2$ GeV$^2$, to get the the Regge trajectories, the mass of scalar glueball $0^{++}$ is lower than those found in table \ref{t2}, but the Regge trajectory related to the pomeron is fine, if you compare with \cite{Landshoff:2001pp}. One can wonder if this scalar glueball state can be related with lowest ``exotic" scalar mode as pointed out in \cite{Brunner:2015oqa}.

The modified softwall model, in the sense used in this work, i.e., solving the problem analytically, is faster than numerical approach and provides satisfactory results. As a further work, we will analyze the complete solution of the problem, i.e., solving the problem numerically to see if there will be any corrections in the results.



\begin{acknowledgments}
 The authors, particularly, EFC would like to thank Danning Li and Frederic Br\"unner  for interesting discussions.
 The authors are partially supported by CAPES, CNPq and FAPERJ, Brazilian agencies.
 
\end{acknowledgments}



\section{Appendix A: Einstein frame and equations of motion}

It is easier to solve the equations of motion in the Einstein frame, which can be defined as, with respect to the string frame: 
\begin{equation}\label{weyl}
 g^E_{mn} = g^s_{mn}e^{-\frac{2}{3}\Phi}, \qquad V^E_G = e^{\frac{4}{3}\Phi}V^s_G\,,
\end{equation}
\begin{equation}\label{redef}
b_E (z) = b_s(z)e^{-\frac{2}{3}\Phi(z)} = e^{A_E(z)}, \qquad A_E(z) = A_s(z) - \frac{2}{3}\Phi(z)\,.
\end{equation}
Then, from action (\ref{acao_corda}) one can obtain the following set of coupled equations (see \cite{Li:2013oda} for more details):
\begin{equation}\label{eq_mov_e_2_1}
 -A''_E + A'^2_E - \frac{4}{9}\Phi'^2  = 0\,,
\end{equation}
\noindent and
\begin{equation}\label{eq_mov_e_2_2}
 \Phi'' + 3A'_E \Phi' - \frac{3}{8}e^{2A_E}\partial_\Phi V^E_G(\Phi) = 0\,.
\end{equation}

\noindent Solving this set of coupled differential equations with the quadratic dilaton background given by (\ref{phi}) one  finds the solutions:
\begin{equation}\label{sol_eq_mov_e_2_1n}
 A_E(z) = \log{\left( \frac{R}{z} \right)} - \log{(_0F_1(5/4, \frac{\Phi^2}{9}))}
\end{equation}
\noindent and
\begin{equation}\label{sol_eq_mov_e_2_2n}
 V^E_G(\Phi) = -\frac{12 ~ _0F_1(1/4, \frac{\Phi^2}{9})^2}{R^2} + \frac{16 ~ _0F_1(5/4, \frac{\Phi^2}{9})^2 \Phi^2}{3 R^2}
\end{equation}

One can use (\ref{weyl}) and (\ref{redef}) to recover the string frame expressions for $A_s(z)$ and $V^s_G(\Phi)$, given by (\ref{redef_2}) and (\ref{vs}), respectively, valid for the dynamical softwall model. 

For our purpose, in the phenomenological modified softwall model, in order to keep this model analytically solvable, we replace $A_s(z)$ showed in (\ref{redef_2}) by $A^s_M(z)$ in (\ref{am}). Consequently the the potential $V^s_G(\Phi)$ for the modified softwall model is now is given by $V^s_M(\Phi) = \exp\{-\frac 43 \Phi\} 16 \Phi^2 /3 R^2$.

As a last comment, $A_E(z)$, $ V^E_G(\Phi)$ and $\Phi(z)$ have to satisfy the eqs. (\ref{eq_mov_e_2_1}) and (\ref{eq_mov_e_2_2}). As we are using an approximation, in order to get an analytical model, it should be clear that $A^s_M(z)$, in fact, does not satisfy exactly the eq. (\ref{eq_mov_e_2_1}).


\begin{thebibliography}{99}
\bibitem{comment}
Actually, we could retain terms up to second order in $k$ without breaking exact solvability, but this contribution would not not modify significantly our subsequent analysis.
\end{thebibliography}
\end{document}